\documentclass[prd,twocolumn,nofootinbib,showpacs]{revtex4}

\def\beq{\begin{equation}} \def\eeq{\end{equation}}
\def\bea{\begin{eqnarray}} \def\eea{\end{eqnarray}}

\def\d{{\rm d}}

\preprint{\vbox{\baselineskip=12pt \rightline{gr-qc/0402064}
\rightline{ICN-UNAM-04/01}}}

\begin{document}

\title{Comments on area spectra in Loop Quantum Gravity}
\author{Alejandro Corichi}\email{corichi@nuclecu.unam.mx}
\affiliation{Instituto de Ciencias Nucleares,\\
Universidad Nacional Aut\'onoma de M\'exico\\
A. Postal 70-543, M\'exico D.F. 04510, M\'exico\\ \ }


\begin{abstract}
We examine and compare different area spectra that have been
recently considered in Loop Quantum Gravity (LQG). In particular
we focus our attention on a Equally Spaced (ES) spectrum operator
introduced by Alekseev {\it et al} that has gained recent
attention. We show that such operator is not well defined within
the LQG framework, and comment on the issue regarding area spectra
and QNM frequencies.
\end{abstract}
 \maketitle

\section{Introduction}

Loop quantum gravity (LQG) has become in the past years a serious
candidate for a non-perturbative quantum theory of gravity
\cite{LQG}. Its most notable predictions are the quantization of
geometry \cite{quantgeom} and the computation of black hole
entropy \cite{abck}. One of its shortcomings is the existence of a
one parameter family of inequivalent quantum theories labelled by
the Immirzi parameter $\gamma$ \cite{immirzi}. The black hole
entropy calculation was proposed as a way of fixing the Immirzi
parameter $\gamma$ (and thus the spectrum of the geometric
operators) when a systematic approach to quantum black hole
entropy was available \cite{abck}. This was used to fix the value
of the Immirzi parameter to the value $\gamma_{\rm
abck}=\frac{\ln(2)}{\pi\sqrt{3}}$ \cite{abck}. Recently, Dreyer
made the suggestion that there is an independent way of fixing the
Immirzi parameter \cite{dreyer}. The new approach is based on a
conjecture by Hod that the {\it real} part of the quasinormal mode
frequencies $\omega_{\rm QNM}$, for large $n$ have an asymptotic
behavior given by \cite{hod} $M\omega_{\rm QNM}=\frac{\ln
3}{8\pi}$. This conjecture was proved analytically by Motl
{\cite{motl}. These modes have an imaginary part that goes to
infinity as $n$ grows, therefore, these are highly damped
oscillatory modes. The conjecture of Hod for the limit of the real
part of the frequency was within the (quantum) framework pioneered
by Bekenstein in which the area spectrum is assumed to be equally
spaced \cite{beken}. Dreyer showed that in order to have
consistence between the BH entropy calculation and QNM
frequencies, one had to assume that the minimum value of $j$ of
the spin network piercing the horizon and contributing
significantly to the entropy had to be $j=1$. With this choice,
the resulting Immirzi parameter would be given by $\gamma_{\rm
d}=\frac{\ln(3)}{2\pi\sqrt{2}}$. He suggested that if the gauge
group of the theory were changed from $SU(2)$ to $SO(3)$ then this
requirement would be immediately satisfied. After that observation
there have been several attempt to suggest different scenarios.
One could classify these attempts in two categories: those that
try to explain the $j=1$ appearance by means of extra requirements
\cite{su(2)} but without changing the geometric operator and those
attempts that suggest modifying the area spectra
\cite{APS,poly,gour}.

In this note we shall focus our attention to this later proposal,
where the Equally-Spaced (ES) area operator is employed
\cite{APS,poly,gour}: \beq \hat{A}(S)_{\rm ES}\cdot\Psi=8\pi
l^2_{P}\,\gamma\sum_{v}(j_v+1/2)\,\Psi \label{esoperator} \eeq
This is to be contrasted to the standard Rovelli-Smolin spectrum
(for simple intersections), \beq  \hat{A}(S)_{\rm
RS}\cdot\Psi=8\pi l^2_{P}\,\gamma\sum_{v}\sqrt{j_v(j_v+1)}\,\Psi
 \eeq
  The ES spectrum has been argued to be relevant for explaining
the $j=1$ contribution while keeping $SU(2)$ as the gauge group
\cite{poly,gour}. Probably the most important property of the
ES-area operator (\ref{esoperator}) is that it assigns a quantum
of area ($4\pi l^2_P\gamma$)  to all edges that pierce the surface
and carry a $j=0$ label.

In this note we shall consider the operator (\ref{esoperator}),
within the framework of LQG. We shall give two different (but
related) arguments to show that the operator is not well defined
in the theory. The first argument will use the `old' language of
Wilson loops and the second argument uses spin networks and
graphs. As we will show, the fact that the operator assigns area
to $j=0$ label is what makes it sense-less. We hope that this note
will help to settle the issue of this particular operator (ar any
operator that `sees' $j=0$ edges for that matter).

The structure of this note is as follows. In Sec.~\ref{sec:2} we
consider the $C^*$-holonomy algebra to show that the ES-operator
does not respect the (Hoop) equivalence classes. In
Sec.~\ref{sec:3} we consider the operator within the graph
perspective and show that cylindrical consistency is violated by
the operator. In Sec.~\ref{sec:4} we comment on the result. Loop
Quantum Gravity experts may safely skip the remainder of the note.

\section{Holonomy Algebras}
\label{sec:2}

In this section we shall analyze the ES area operator as seen from
the perspective of Holonomy Algebras (HA) and the GNS construction
of the (kinematical) Hilbert space ${\cal H}=L^2(\overline{{\cal
A}/{\cal G}},\d \mu_{\rm AL})$ of the theory. It is now well
understood that there are several ways of characterizing the
quantum configuration space $\overline{{\cal A}/{\cal G}}$ of
gauge invariant generalized connections and of the Hilbert space.
Historically, the first construction made use of the fact that one
could define an Abelian $C^*$-algebra of configuration
observables, the so-called {\it Holonomy Algebra} ${\cal HA}$
\cite{aa:ci}. The Gelfand-Naimark theory tells us that the ${\cal
HA}$ can be seen as the space  of continuous functions $C(\Delta)$
on the spectrum $\Delta$ of the algebra ${\cal HA}$. This is the
quantum configurations space. Furthermore, the Hilbert space can
be constructed via the GNS procedure for a properly defined
positive functional (the so called Ashtekar-Lewandowski state).
Since the elements of the Hilbert space are to be built out of
elements $h_\alpha\in {\cal HA}$ of the holonomy algebra, then it
better be that any operator $\hat{\cal O}$ in ${\cal H}$ respects
the algebraic properties of the algebra ${\cal HA}$ if it is to be
well defined.

Why should there be any problem? The reason for the existence of
consistency conditions to be met is that the elements of ${\cal
HA}$ are {\it equivalence classes} of loops (closely related to
Wilson loops), where two loops $\alpha$ and $\beta$ are equivalent
if the holonomies along them are the same {\it for all}
connections. Furthermore, in order to define ${\cal HA}$, one
needs to quotient the original algebra by an ideal that takes care
of the so-called Mandelstam identities arriving at a new
equivalence class $K$ (for details see \cite{aa:ci}).

Thus, there are loops that are K-equivalent to the zero loop, and
the corresponding algebra element $[h_{\alpha=0}]={\rm Id}$,
correspond to the unit element. The unit element of the algebra,
as its name indicates, can be multiplied freely and the resulting
state is the same in the GNS construction.

Now, how do we make contact with the operator (\ref{esoperator})
and the $j=0$ spin networks? A closed loop is a particular case of
a closed graph, and we can define a spin network there by
assigning representation of $SU(2)$ to it, labelled by $j$. If one
chooses $j=0$, one has the trivial (identity) function, and
therefore the unit element $h_{\alpha=0}$. That is, The zero-$j$
spin networks correspond to an element of the algebra equivalent
to the zero-loop, or, in other words the unit element of the
algebra. This means that we can add ar remove closed loops with
zero-$j$ for free to a state and get the ``same physical state".
The ES-area operator (\ref{esoperator}) endows with different
eigen-values for the area to the state each time one adds or
removes a $j=0$ loop that crosses the surface. Thus, the fact that
the ES operator counts zero-$j$ spin networks and assigns area to
them means that its action depends on the representative of the
equivalence class $[h_{\alpha}]$. The operator does not respect
the K-equivalence classes and is, therefore, not well defined on
the Hilbert space ${\cal H}$ of the theory.

\section{Graphs}
\label{sec:3}

There are alternative ways of characterizing the Hilbert space
${\cal H}$ and the quantum configuration space $\overline{{\cal
A}/{\cal G}}$ of the theory. Of particular relevance are the
so-called projective techniques that make use of families of
graphs and projective families (for a nice review see
\cite{aa:jl2}). The basic idea is to define a family of quantum
theories that live on closed graphs $\Gamma$, corresponding
roughly speaking to a latice gauge theory on the graph. The
continuum theory is recovered by taking the projective limit of
the {\it largest} graph.

To be concrete, if we have a graph $\Gamma$ and a spin network
$\Psi_{(\Gamma,\vec{j})}(A)$ on it, we can define a unique
function $\Psi^\prime_{(\Gamma^\prime,\vec{j})}(A)$ on a larger
graph $\Gamma^\prime > \Gamma$ as follows: If $\Gamma^\prime >
\Gamma$ is such that can be obtained by $\Gamma$ by adding
artificial vertices to already existing edges, define the new
function by trivial composition \cite{aa:jl2}.
 If the graph $\Gamma^\prime
> \Gamma$ contains new edges, then the new function
$\Psi^\prime_{(\Gamma^\prime,\vec{j})}(A)$ is obtained by
assigning the identity function to each new edge. This means
defining a new spin network with $j_I=0$ {\it for all new edges}
$e_I$.

Thus, for each spin network on $\Gamma$, there exist an infinity
of spin network  states defined on any larger graph $\Gamma^\prime
> \Gamma$, with lots of $j=0$ edges. A function on the full
Hilbert space is made of the collection of all these functions
that are part of the `cylindrical family'.

Any operator $\hat{\cal O}$ of the full theory needs to satisfy
what is called {\it cylindrical consistency}, which means that its
action should be the same for all elements of the family. Now, we
come back to the $j=0$ spin networks. If the operator $\hat{\cal
O}_\Gamma$ is able to see the $j=0$ edges of the graph $\Gamma$,
then its action will depend on the element of the cylindrical
family and therefore will not be consistent. We can then state
that any operator that acts non-trivially on a given graph on
$j=0$ edges, will not be part of a consistent cylindrical family
of operators, and will {\it not} define an operator on the
continuum. The ES-area operator (\ref{esoperator}) is clearly an
example of this class of operators and is, therefore,  not well
defined.

As can be expected, the reason why the operator does not exist is
simple to understand and can be seem from these two (slightly
different) perspectives. In fact, the langauge of loops or closed
graphs is only a matter of convenience but they are equivalent.
Every graph $\Gamma$ can be decomposed into $N$ independent loops
$\alpha_i,\,i=1,\ldots, N$. On the other hand, the graph might
have $M$ edges $e_I,\,I=1,\ldots,M$, with $M\geq N$, and therefore
any cylindrical function is a function $f_\Gamma: G^M \mapsto C$
from $M$ copies of the gauge group $G$ to the complex numbers. On
the other hand one has $N$ Wilson loops
$W[\alpha_i,A]=\frac{1}{2}{\rm Tr}{\cal P}\exp(\oint_{\alpha_i}
A)$ that are complex valued functions. The statement is that any
Spin network $\Psi_{(\Gamma,\vec{j},\vec{m})}(A)$ on $\Gamma$ can
be written as a polynomial of degree given by the maximum value of
the labels $j_I$ as follows,
$$\Psi_{(\Gamma,\vec{j},\vec{m})}(A)=\sum_{n_i}A_{n_1\cdots n_N}
W[\alpha_1]^{n_1}W[\alpha_2]^{n_2}\cdots W[\alpha_N]^{n_N}
$$
The advantage of working with spin networks is that they form a
convenient basis that diagonalizes the geometric operators, in
particular, the area operator for simple intersections of the spin
network and the surface $S$.

\section{Discussion}
\label{sec:4}

In the previous sections we have shown that the ES-area operator
as proposed by Alekseev, Polychronakos and Smedback (APS)
\cite{APS}, and used in Refs. \cite{poly,gour}, is not a valid
operator in LQG from the mathematical viewpoint, using arguments
in both the GNS construction and in the projective families
construction. This conclusion also applies to the length operator
recently suggested in $2+1$ gravity in \cite{Freidel:2002hx}.

It has been noted that one might modify the operator
(\ref{esoperator}) such that it {\it is} well defined, by changing
its action when acting on a $j=0$ edge. The choice that makes it
well defined is to ask that the new operator $\hat{A}^\prime(S)$
annihilates the state (i.e. it yields zero eigenvalue). This is
not the action that was originally proposed by APS \cite{APS} (and
 analyzed later on by Polychronakos in Ref.~\cite{poly}), where
the operator was motivated by a new regularization that included
`quantum corrections', with a resulting behavior similar to the
zero point energy of a harmonic oscillator \cite{APS}. With this
modification, the new and well defined operator
$\hat{A}^\prime(S)$ would cease to be Equally-Spaced (ES), since
there would be a larger area gap from $j=0$ to $j=1/2$ edges of
$8\pi l^2_P\gamma$, as opposed to $4\pi l^2_P\gamma$ that is the
the area gap in the rest of the ES part of the spectrum. This
would presumably make it less appealing for providing an
explanation of the QNM frequencies.

There might be some further considerations on why an ES-area
spectrum (without the nontrivial contribution from $j=0$) is not
the most desirable one, such as the so-called Bekenstein-Mukhanov
effect \cite{lee,dreyer,motl}, but we shall not go further into
that discussion (see, for instance, the first reference of
\cite{LQG} and \cite{poly} for some discussion).

The standard spectrum of Rovelli-Smolin, not only has been
obtained by different regularization procedures \cite{quantgeom},
but seems to be robust given its physical and mathematical
properties. However, whether or not Loop Quantum Gravity (with the
Rovelli-Smolin spectrum) should have anything to say about the
asymptotic Quasi-Normal Modes frequencies remains, in our opinion,
an open issue. The reason for this is that recent numerical and
analytical explorations of charged and rotating Black Holes do not
show the asymptotic behavior that one would expect if one assumes
a Bohr correspondence principle, as originally conjectured by Hod
\cite{hod} (for an incomplete list of recent references in QNM see
\cite{qnm:new}).

Finally, let us note that a similar argument to that presented in
Sec.~\ref{sec:3} has already been given in \cite{zapata}, from a
slightly different perspective.

\begin{acknowledgments}
The author would like to thank C. Fleischhack for comments. This
work was partially supported by a DGAPA-UNAM grant No. IN112401
and a CONACyT grant No. J32754-E.
\end{acknowledgments}

\end{document}